\documentclass[conference]{IEEEtran}
\usepackage[latin9]{inputenc}
\usepackage{amsmath}
\usepackage{amssymb}
\usepackage{graphicx}

\makeatletter

\DeclareTextSymbolDefault{\textquotedbl}{T1}
\providecommand{\tabularnewline}{\\}

\IEEEoverridecommandlockouts
\usepackage{cite}
\usepackage{amsfonts}\usepackage{algorithmic}
\usepackage{textcomp}
\usepackage{xcolor}
\def\BibTeX{{\rm B\kern-.05em{\sc i\kern-.025em b}\kern-.08em
    T\kern-.1667em\lower.7ex\hbox{E}\kern-.125emX}}

\makeatother

\begin{document}
\title{Multi-Component V2X Applications Placement in Edge Computing Environment}
\author{\IEEEauthorblockN{Ibrahim Shaer\textsuperscript{{*}}, Anwar Haque\textsuperscript{$\dagger$},
and Abdallah Shami\textsuperscript{{*}}}\IEEEauthorblockA{\textsuperscript{{*}}Department of Computer and Electrical Engineering,\textsuperscript{$\dagger$}Department
of Computer Science\\
Western University, London ON, Canada\\
\{ishaer, ahaque32, abdallah.shami\}@uwo.ca}}
\maketitle
\begin{abstract}
Vehicle-to-everything (V2X) services are attracting a lot of attention
in the research and industry communities due to their applicability
in the landscape of connected and autonomous vehicles. Such applications
have stringent performance requirements in terms of complex data processing
and low latency communications which are utilized to ensure road safety
and improve road conditions. To address these challenges, the placement
of V2X applications through leveraging of edge computing paradigm,
that distributes the computing capabilities to access points in proximity
to the vehicles, presents itself as a viable solution. However, the
realistic implementation of the edge enabled V2X applications is hindered
by the limited computational power provided at the edge and the nature
of V2X applications that are composed of multiple independent V2X
basic services. To address these challenges, this work targets the
efficient placement of V2X basic services in a highway scenario subject
to the delay constraints of V2X applications using them and the limited
computational resources at the edge. To that end, this work formulates
a binary integer linear programming model that minimizes the delay
of V2X applications while satisfying the resource requirements of
V2X basic services. To demonstrate the soundness of the approach,
simulations with varying vehicle densities were conducted, and the
results reported show that it can satisfy the delay requirements of
V2X applications.
\end{abstract}

\begin{IEEEkeywords}
V2X applications, edge computing, placement, multi-component
\end{IEEEkeywords}

\section{Introduction}

Intelligent Transportation Systems (ITSs) are envisioned to ameliorate
traffic congestion and improve road safety and traffic experience.
ITSs have drawn the attention of a large number of stakeholders due
to their direct effect on the manufacturing of sensor and wireless-equipped
vehicles known as connected and autonomous cars. In this regard, Vehicle-to-everything
(V2X) applications are considered a key enabler for the shift to Intelligent
Transportation Systems (ITSs) in terms of traffic management. These
applications allow the vehicles to communicate and exchange information
with their surrounding environment that includes other vehicles, pedestrians
and supporting road side units (RSUs). To ensure road safety, these
applications operate with stringent end-to-end (E2E) latency/delay.
There are different paradigms that can determine the placement of
V2X applications to address the E2E latency requirements. The placement
of these services is disruptive to the customary cloud-based infrastructure.
The projected increase in the number of connected and autonomous vehicles
will result in data explosion. The data will be routed to a single
centralized server creating severe network traffic congestion \cite{b1}
. Additionally, the centralized servers are usually located far from
vehicles generating data; thus, incurring a huge E2E delay. Furthermore,
this architecture exposes a single point of failure, which is huge
risk to take for time and mission-critical V2X applications. Given
these circumstances, distributing the cloud computing technology in
proximity to users is proposed as a viable solution to deal with the
shortcomings of the centralized paradigm \cite{b2}. This computing
architecture is referred to as Edge Computing. Edge Computing can
support the latency requirements of the V2X applications which are
critical for their performance \cite{b3}. In addition, the edge servers
collect data from the close local nodes which allows for a more individualized
experience for the V2X application users. While Edge Computing paradigm
can ensure some V2X system-level performances, this comes at the expense
of limited computational power at the edges which hinders the processing
of large amount of data. Microservices architecture, that decomposes
a single application into decoupled modules, combined with virtualization
techniques, that fully utilizes resource at the edges, can be used
to address this issue. Hawilo \textit{et. al }\cite{key-3}\textit{
}investigated the applicability of this paradigm for Virtual Network
Functions which display similar characteristics to V2X applications
making it a viable option for their placement. In the domain of V2X
applications, 3rd Generation Partnership Project (3GPP) \cite{b4}
envisions complex V2X applications that combine vehicle status analysis,
imminent traffic events generation, and raw sensor data exchange that
define the function of autonomous and connected vehicles. Each of
these applications rely on the data processing and analysis of miniscule
V2X basic services.

Mobile edge clouds (MEC), edge clouds and roadside cloud have been
proposed in several previous works in the context of vehicular applications.
In \cite{b5}, Emara \textit{et. al} employ an MEC-assisted architecture
to evaluate end-to-end latency for vehicles to detect the vulnerable
road units. Moubayed \textit{et. al }\cite{b6}\textit{ }formulated
an integer linear programming problem for efficient placement of V2X
basic service taking into consideration V2X basic services' delay
and computational requirements in a hybrid environment that includes
edge and core nodes. Supporting V2X applications while considering
the vehicle\textquoteright s mobility aspects has been extensively
addressed in literature. To support V2X applications, \cite{b7,b8,b9}
consider migrating the services according to vehicle\textquoteright s
mobility. In \cite{b7}, the authors customize a three-layered architecture
that consists of a vehicular cloud, a roadside cloud and a central
cloud to support vehicular applications. Their approach focuses on
the dynamic allocation of resources, driven by the vehicle\textquoteright s
mobility, in vehicular and roadside clouds. In \cite{b8}, Yu \textit{et.
al} consider the migration of V2X applications placed on edge servers
according to predictive vehicle\textquoteright s mobility combined
with setting a priority schema for V2X applications. The approach
considers the latency and resource requirements of each of the applications.
In the same context, Yao \textit{et. al }\cite{b9} investigate Virtual
Machine (VM) placement and migration in roadside cloud that is part
of the vehicular cloud computing architecture. The approach targets
minimizing the overall network cost given the available resources
at the edge. Each of these previous works has its own shortcoming.
One common aspect is considering either latency or resource limitations
for the placement of the services at the edge. Another shortcoming
is the disregard of the nature of V2X applications that may be composed
of a single or many modules. Finally, different traffic conditions
were not considered to model any solution for vehicular application
placement. To address these shortcomings, this work focuses on V2X
application placement that minimizes the end-to-end delay which takes
into consideration the computational requirements of V2X services
forming it. This work\textquoteright s main contributions are as follows:
\begin{itemize}
\item Decompose V2X applications into multi-V2X basic services.
\item Formulate the optimal V2X application placement by considering their
delay requirements and the resource requirements of their constituent
components.
\item Evaluate the performance of the optimal placement in terms of average
delay and density distribution for each V2X application under different
traffic conditions.
\end{itemize}
The remainder of this paper is organized as follows: Section II describes
the system model and presents the problem formulation, Section III
provides the simulation procedure and discusses the results and Section
IV concludes the paper and suggests future work.

\section{System Model}

In the reference model, a highway scenario is considered. Each of
the vehicles moving on the highway is running a set of V2X applications
that are collecting data from nearby roadside units (RSUs) to function
autonomously. RSUs and the vehicles are communicating directly using
Dedicated Short-range Communication \cite{b10}, and no communication
takes place between any vehicles. Each RSU is equipped with a server
which are both considered as an edge computing node. Vehicles are
receiving data from V2X basic services placed on each RSU. European
Telecommunication Standardization Institute (ETSI) defines three V2X
basic services that are the foundation of any envisioned V2X applications.
The V2X basic services are as follows: Cooperative Awareness Basic
Service (CA) \cite{b11} is responsible for creating, analyzing and
sending Cooperative Awareness Basic Messages (CAMs) which include
information about the vehicle\textquoteright s status and attributes,
Decentralized Environmental Notification Service \cite{b12} (DEN)
broadcasts Decentralized Notification Messages (DENM) whenever a road
hazard or abnormal traffic condition takes place, and Media Downloading
\cite{b13} service is requested on demand by the passengers of the
vehicle. Additionally, ETSI defines Local Dynamic Maps (LDMs) \cite{b14}
that are responsible for storing the sent CAMs and DENMs. Because
LDMs store spatial relevant information, an LDM is deployed on each
edge server. LDMs are queried by V2X basic services in order to retrieve
information. Finally, in addition to basic vehicular services that
are related to road safety, there is a variety of innovative applications
that are referred to as value-added services that are of lower priority
\cite{b15}. These services include augmented reality, parking location
and others that are part of the infotainment services provided by
vehicular applications. Compared to road safety applications, these
services display high levels of diversity and individuation. Therefore,
they need to be migrated when the vehicle moves from one edge server
coverage zone to another. For this purpose, each edge server reserves
part of its resources to accommodate these migrating services. In
this section, the system design and the optimal optimization technique
for V2X basic service placement are presented.

\subsection{System Design}

In the reference model used for the placement of V2X basic services,
HWY 416 IC-721A that passes through the city of Ottawa is considered.
The edge computing servers are deployed uniformly along the highway
as the deployment of RSU is out of the scope of this paper. No communication
interference zone exists between any two successive RSUs to avoid
the possibility of encountering ping-pong handover cases which will
be difficult to handle in an optimization model. Additionally, the
vehicles are assumed to be always connected to RSUs throughout their
journey. The end-to-end latency of a service is the sum of the communication,
processing, transmission and propagation delay. The propagation delay
is dependent on the medium of communication which is out of the scope
of this paper, and therefore considered negligible. In this model,
DSRC, the communication technology between the moving vehicles and
RSUs, affects the communication delay. In the proposed model, the
processing and transmission delay between the communicated edge servers
is considered. Each edge computing server has the same computing and
processing power that are expressed by the number of cores and RAM
available. Finally, the vehicle density is considered to model real
case scenarios.

\subsection{Optimization Problem}

In the optimization function, a set of edge servers and V2X services
are considered. Let $N$ denote the set of edge servers where $n\in N.$
Let $U$ denote the set of unique V2X basic services where $u\in U.$
The availability of the computational resources on the edge is denoted
by matrix $Cap$ where $Cap_{kn}$ denotes the $k$th computational
resources available on edge server $n$. Matrix $R$ represents the
resources required by the V2X basic services where $R_{ku}$ represents
the $k$th computational resources required by V2X basic service $u.$
A binary row vector $\overrightarrow{q}$ denotes the edge servers
a vehicle can communicate with. Let $C$ be the matrix that represents
the processing and the transmission latency between edge servers where
$C_{ij}$ represents the latency between edge server $i$ and edge
server $j$. Matrix $M$ represents the V2X services needed by V2X
applications where $M_{au}=1$ denotes that application $a$ needs
V2X basic service $u$. Let $X$ denote the placement matrix where
$X_{un}=1$ means that V2X basic service $u$ is placed on edge server
$n$. The column $X_{u}$ denotes the placement of the V2X services
on edge server $n$. $D_{a}^{v}$ and $D_{a}^{th}$ denote respectively
the delay experienced by a moving vehicle $v$ served by application
$a$ and the maximum tolerable threshold of this delay. To represent
the vehicles' density, $\gamma$ is used. $d_{com}^{v}$ and $d_{DL}^{v}$
denote respectively the communication and download latency between
a vehicle $v$ and a serving edge server. The optimization function
used to minimize the delay of V2X application is as follows:

\begin{equation}
\mathop{min\sum_{a\in A}}D_{a}^{v}
\end{equation}

where:

\begin{equation}
D_{a}^{v}=d_{com}^{v}+max(M_{a}\odot min(X\odot(\gamma C\times\overrightarrow{q}))+d_{DL}^{v}
\end{equation}

subject to:

\begin{equation}
D_{a}^{v}\leq D_{a}^{th},\forall a\in A
\end{equation}

\begin{equation}
RX\leq Cap
\end{equation}

\begin{equation}
\sum X_{n}=1,\forall n\in N
\end{equation}

In what follows, the explanation of the equations (1)-(5). Equation
(1) describes the overall objective which is minimizing the summation
of the delay of all V2X application experienced by a vehicle requesting
their services. Equation (2) presents the components contributing
to the delay of V2X application. The delay of a V2X application is
the delay of the V2X services it relies on depending on the edge servers
a vehicle can communicate with. Because the functioning of a V2X basic
service is independent of other V2X basic services, the delay of a
V2X application is defined as the maximum of the delay of its constituent
V2X basic services. This value is added to the communication and download
link delay. Next, the equations (3) -- (5) describe the constraints.
Equation (3) defines that the delay of an application should not exceed
its maximum defined tolerable delay. Equation (4) ensures that the
resources allocated to a V2X service do not exceed the available resource
on the hosting edge server. Equation (5) limits placing only one V2X
service on each edge server. The following diagram illustrates an
example of the communication and processing that takes place for a
V2X application that requires CA and DEN services, given that each
server has resources reserved for migrating applications denoted by
VM 3. 
\begin{figure}
\begin{centering}
\par\end{centering}
\begin{centering}
\caption{System Model}
\par\end{centering}
\centering{}\centerline{\includegraphics[scale=0.4]{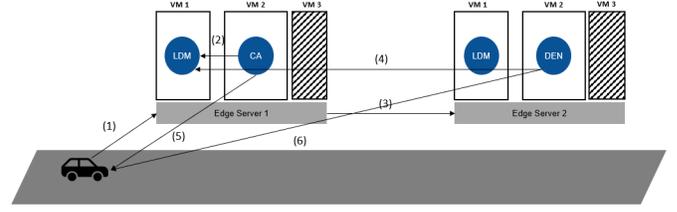}}
\end{figure}
The logic governing the realization of the application is as follows:

(1) The vehicle requests the services of an application. This step
incurs communication delay that is denoted by $d_{com}^{v}.$

(2) The CA service found on Edge Server 1 requests the necessary information
from the LDM. The processing of the request on the LDM is denoted
by $C_{11}$.

(3) Edge server 1 communicates with the closest server that include
DEN basic service. No delay is considered in this phase.

(4) DEN queries and receives information from the LDM that is closest
to the requesting vehicle. The LDM on edge server 1 has accurate information
about the requesting vehicle's surrounding environment. This delay
is the sum of the processing delay of LDM on edge server 1 and the
transmission delay between edge server 1 and 2. This is denoted by
$C_{12}.$

(5, 6) These steps represent the CA and DEN response to the requesting
vehicle. This delay is denoted by $d_{DL}^{v}$.

For basic service CA, the delay is as follows:

\[
d_{CA}^{v}=C_{11}+d_{DL}^{v}
\]

Similarly, the delay for DEN is:

\[
d_{DEN}^{v}=C_{12}+d_{DL}^{v}
\]

Given that the requests for each basic service are executed in parallel
and that these services are independent in their execution, the delay
experienced by a vehicle $v$ requesting the services of application
$a$ is:

\[
D_{a}^{v}=d_{com}^{v}+max\{d_{CA}^{v},d_{DEN}^{v}\}
\]

\section{Experimental Setup and Results}

\subsection{Simulation Setup}

In order to evaluate the placement of V2X basic services, a realistic
simulation environment must be created. To this end, Simulation of
Urban Mobility (SUMO) \cite{b16} was used to extract the movement
of vehicles along a highway. A 4 km highway that resembles HWY 416
IC-712A was considered as a reference highway. Ontario traffic volume
for provincial highways \cite{b17} provided the average daily traffic
and the accident rates during summer, winter, weekdays and weekends.
In the simulation setup, the statistics offered by this report were
used to emulate moderate and heavy traffic experienced on HWY 416
IC-712A highway that is expressed through the vehicles per hour parameter
in SUMO. Regarding the movement of the vehicles, Table I summarizes
the key parameters that are used in the simulation.

\begin{table}
\caption{SUMO vehicle movement parameters}

\centering{}%
\begin{tabular}{|c|c|}
\hline 
\textbf{\scriptsize{}Parameter} & \textbf{\scriptsize{}Value}\tabularnewline
\hline 
\hline 
{\scriptsize{}Maximum Speed} & \textbf{\scriptsize{}$27.7m/s$}\tabularnewline
\hline 
{\scriptsize{}Maximum Acceleration} & \textbf{\scriptsize{}$2.6m/s^{2}$}\tabularnewline
\hline 
{\scriptsize{}Maximum Deceleration} & \textbf{\scriptsize{}$4.5m/s^{2}$}\tabularnewline
\hline 
\end{tabular}
\end{table}

The V2X applications considered are Platooning (PL), Sensor and Sensor
State Mapping (SSM), Emergency Stop (ES), Pre-crash Sensing Warning
(PSW) and Forward Collision Warning (FCW). Their corresponding performance
requirements and service components are presented in Table II \cite{b18,b19}.
Choosing these V2X applications stems from their importance and stringent
performance requirements in the realm of the autonomous cars. In addition,
in the context of the defined problem, each of the chosen V2X application
offers a unique combination of V2X services. In the simulation procedure,
the communication delay between a vehicle and an RSU is 1 ms \cite{b20}.
In this model, the processing delay is the amount of time required
by a Local Dynamic Map to process the data requested by other V2X
services either placed on the same or different edge server. In \cite{b21},
the authors devise an LDM according to the specifications defined
by ETSI. The application defines two Application Programming Interfaces
(APIs) that retrieve information of the IDs of the vehicles driving
on the same road and the vehicle driving immediately ahead of the
requesting vehicle. For different number of queried vehicles ranging
from 5 to 20 vehicles, the response time was between 3 and 5 ms with
no clear correlation between the size of the data and the response
time. Consequently, in the simulation setup, the processing delay
is generated uniformly between 3 and 5 ms. In the same context, the
authors in \cite{b22}, assumed the transmission latency between two
edge servers to be between 1 and 5 ms. Because the simulation procedure
takes place under several vehicle densities, the increase of the data
processing and transmission overhead with the increase of number of
vehicles is inevitable. In this regard, the execution cost increases
with the number of vehicles in proximity to the vehicle requesting
the V2X application services. As the implementation of LDM did not
consider cases beyond 20 vehicles, the added delay for these cases
will be in the form of $\log(NC/20)$ where $NC$ represents the number
of cars and the expression is derived from the increase of processing
delay upon the increase in size of the queried data in SQL \cite{b23}.
In terms of edge servers, 10 edge servers are deployed every 400 m
alongside the highway. Each of the RSUs hosts an LDM, V2X service
and an optional migrating service. The computational requirements
of CA, DEN and Media services are those of a small, medium and large
VMs. Table III summarizes the edge server capabilities and the computational
requirements of CA, DEN and Media services. In the experimental procedure,
the placement of the V2X basic services is carried out using the defined
optimization function. Next, traffic simulation is executed for defined
densities that reflect moderate and heavy traffic. The traffic traces
were generated for 1500 seconds. Every 10 seconds, a snapshot of the
road condition is taken and delays for each V2X application for each
vehicle is calculated. Finally, at the end of the simulation, the
average delay for each V2X application is obtained.

\begin{table}
\caption{Breakdown of V2X applications' V2X basic service and performance metrics}

\centering{}{\scriptsize{}}%
\begin{tabular}{|c|c|c|c|}
\hline 
\textbf{\scriptsize{}Application} & \textbf{\scriptsize{}Service(s)} & \textbf{\scriptsize{}Latency(ms)} & \textbf{\scriptsize{}Reliability(\%)}\tabularnewline
\hline 
\hline 
{\scriptsize{}PL} & {\scriptsize{}CA} & {\scriptsize{}50} & {\scriptsize{}90}\tabularnewline
\hline 
{\scriptsize{}SSM} & {\scriptsize{}CA, DEN, Media} & {\scriptsize{}20} & {\scriptsize{}90}\tabularnewline
\hline 
{\scriptsize{}ES} & {\scriptsize{}DEN} & {\scriptsize{}10} & {\scriptsize{}95}\tabularnewline
\hline 
{\scriptsize{}PSW} & {\scriptsize{}CA, DEN} & {\scriptsize{}20} & {\scriptsize{}95}\tabularnewline
\hline 
{\scriptsize{}FCW} & {\scriptsize{}CA, DEN} & {\scriptsize{}10} & {\scriptsize{}95}\tabularnewline
\hline 
\end{tabular}{\scriptsize\par}
\end{table}

\begin{table}

\begin{centering}
\caption{Computational Requirements}
\par\end{centering}
\begin{centering}
\begin{tabular}{|c|c|c|}
\hline 
\textbf{\scriptsize{}Entity} & \textbf{\scriptsize{}Number of Cores} & \textbf{\scriptsize{}RAM}\tabularnewline
\hline 
\hline 
{\scriptsize{}Edge Server} & {\scriptsize{}8} & {\scriptsize{}8}\tabularnewline
\hline 
{\scriptsize{}CA} & {\scriptsize{}2} & {\scriptsize{}2}\tabularnewline
\hline 
{\scriptsize{}DEN} & {\scriptsize{}2} & {\scriptsize{}4}\tabularnewline
\hline 
{\scriptsize{}Media Service} & {\scriptsize{}4} & {\scriptsize{}6}\tabularnewline
\hline 
{\scriptsize{}LDM} & {\scriptsize{}4} & {\scriptsize{}2}\tabularnewline
\hline 
\end{tabular}
\par\end{centering}
\end{table}

\subsection{Implementation}

The optimization function was solved using IBM ILOG CPLEX 12.9.0 through
its Python API. The solution is provided instantly for all simulation
scenarios with different vehicle densities on a laptop with an Intel
Core i7-8750 CPU, 2.21 GHz clock frequency and 16 GB of RAM. The final
solution includes the V2X services placed on each edge server.  

\subsection{Results and Discussion}

To evaluate the efficacy of the optimization function, the simulation
procedure was carried out using two different traffic scenarios each
representing moderate (Scenario 1) and heavy (Scenario 2) traffic
models. The results are obtained as an average for five independent
runs. To assess the placement function, the average delay of each
V2X application under study is obtained and compared it to the maximum
tolerable delay. Additionally, the model is evaluated using the probability
density function of each of V2X application. The density function
provides a more thorough overview about the distribution of the delays
in terms of detecting extreme values that are overshadowed by the
common values trend. Furthermore, the density functions reveal the
shortcomings of the approaches that are concealed by the calculation
of the mean. The suggested optimization function failed to converge,
so a new heuristic algorithm that relaxes the delay threshold for
each application by magnitudes of the reliability metrics is considered
and executed. This heuristic algorithm is referred to as: Resource
and Delay-aware V2X basic service Placement (RDP). The results of
the simulation process in terms of the average delay and the probability
densities of each V2X application are presented in Figures 2-5. 

\begin{figure}
\caption{Average Delay of V2X Applications for different Vehicle Densities}

\begin{centering}
\centerline{\includegraphics[scale=0.225]{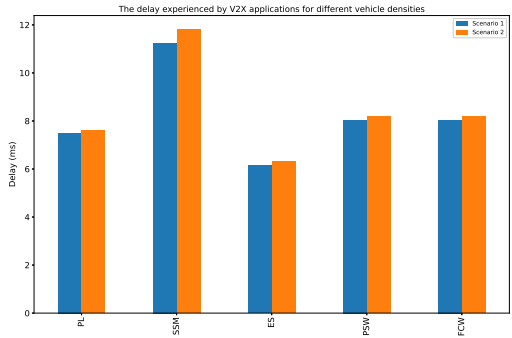}}
\par\end{centering}
\end{figure}

\begin{figure}
\begin{centering}
\par\end{centering}
\begin{centering}
\caption{Probability Density Function for 1500 vehicles/hour}
\par\end{centering}
\centering{}\centerline{\includegraphics[scale=0.13]{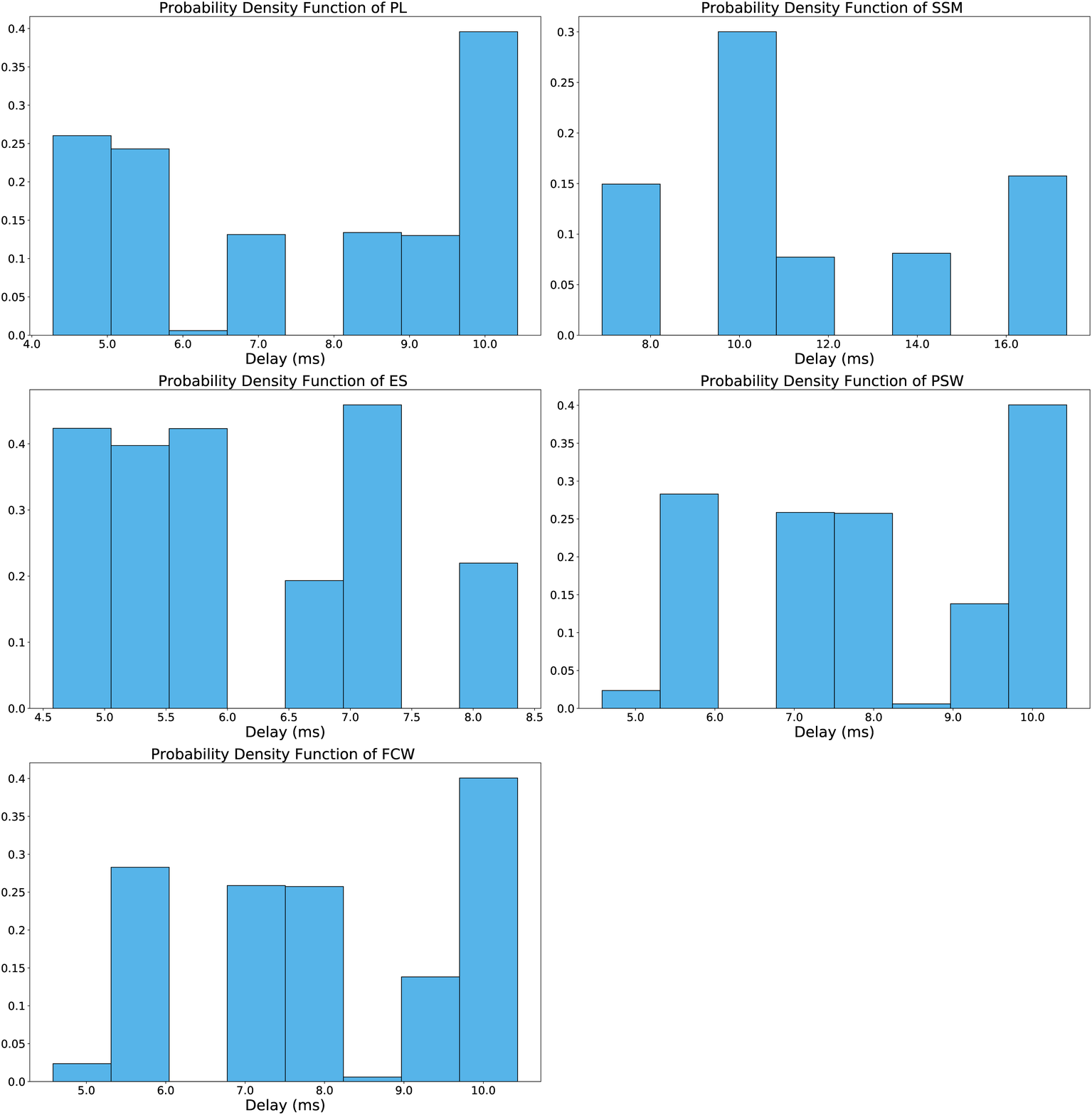}}
\end{figure}

\begin{figure}

\caption{Probability Density Function for 1800 vehicles/hour}

\centering{}\centerline{\includegraphics[scale=0.13]{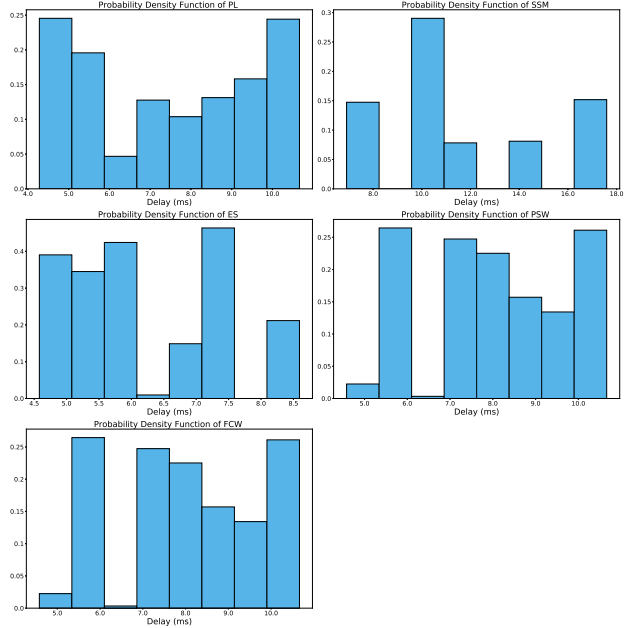}}
\end{figure}
 Figure 2 shows the mean delay for each of the V2X applications. The
results clearly show that the average delay experienced by each V2X
application is within the tolerable threshold. These results show
that the heuristic algorithm met the stringent V2X application delay
requirements. In terms of the traffic effect, the mean of delay of
each application has slightly increased but still fulfills the overall
objective of the placement function. The vehicles' density has contributed
to an increase in the average delay of the V2X applications in the
range of 1.3\% to 4.8\% whereby the SSM application has experienced
the greatest variation. This fact shows that the placement of Media
Service, that SSM relies on, is the most sensitive to traffic variation.
On the other hand, the probability density functions tell a different
story. Figures 3 and 4 depicting the delay distribution for both cases
show that the delay is highly skewed to the left which supports the
viability of the approach. However, this is not the case for FCW application
which shows that for each scenario, 20\% and 25\% of the experienced
delay exceeds the tolerable threshold which is beyond the 5\% permitted
shown in Table II. In terms of traffic effect, it is observed that
there is a slight right shift of the probability distribution in scenario
2. Additionally, it is observed that some applications have similar
probability distributions. This is attributed to the fact that these
applications need the same V2X services, and as it shows, these services
incur the most delay out of the other services that they rely on.
The dispersion of some of the probability density function is due
to the limited number of edge nodes hosting V2X services. The limited
number of edge servers means that vehicles at the start and the end
of the route will suffer from prolonged delay due to the distance
separating the vehicles and the closest V2X basic services. In the
cases of continued route, the suggested approach can be replicated
along the highway to ensure that V2X services are delivered as expected.
For comparison purposes and to further cement this paper\textquoteright s
approach, a baseline approach that maximizes the resource utilization
at each node server is compared to RDP. The baseline approach formulates
a placement algorithm that takes into consideration only the available
resources at each node. This baseline approach is reffered to as Resource-Aware
Algorithm (RAA). The two approaches were evaluated according to the
probability density functions of the delays of ES and FCW applications.
The probability densities are depicted in Figure 5.

\begin{figure}

\begin{centering}
\caption{RDP vs RAA}
\par\end{centering}
\raggedright{}\centerline{\includegraphics[scale=0.21]{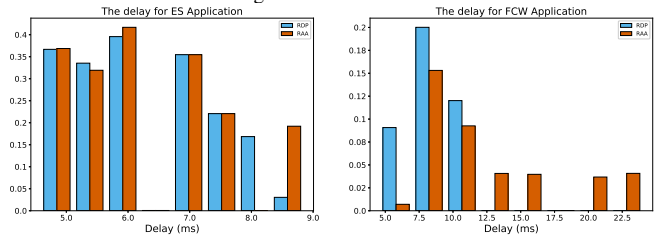}}
\end{figure}

The baseline line approach\textquoteright s density function shows
promising results regarding the ES application as the full delay distribution
is below the tolerable threshold. More values are concentrated on
the extremes which makes it harder to gauge its value whenever the
application is requested. However, for the case of FCW, this approach
fails to be within the tolerable threshold rendering this approach
ineffective for mission-critical applications. This is to be expected
given that FCW application requires CA and DEN basic services. Due
to the nature of RAA that maximizes the overall resource utilization,
deploying more of CA services results in decreasing the utilization
which incurs extra delay for FCW application when requesting the services
of CA.

\section{Conclusion}

This paper addressed the efficient placement of V2X basic service
comprising different V2X applicatins in an edge computing environment.
To this end, an optimization function that minimizes the delay for
multi-component V2X applications consisting of V2X services while
considering the resource requirements of these services under different
traffic conditions is formulated. The approach was evaluated under
realistic scenarios where homogeneous edge servers with limited computational
power and variable traffic conditions were considered. Furthermore,
the approach was compared to a baseline approach that maximizes the
overall resource utilization of edge servers. The results have shown
that the approach guarantees an acceptable quality of service, and
outperforms other approaches while emulating realistic conditions.
While the current work considers that each V2X application has a constant
request rate, the plan is to extend the work to consider different
request distributions to mimic a real-world scenario. In the same
context, the deployment of V2X applications in a dynamic service availability
environment is also a subject to our future work.

\end{document}